\documentstyle[prc,aps,epsfig]{revtex}
\newcommand{\be}{\begin{eqnarray}}
\newcommand{\ee}{\end{eqnarray}}

\newcommand{\bi}{\begin{itemize}}
\newcommand{\ei}{\end{itemize}}
\begin{document}
\twocolumn[\hsize\textwidth\columnwidth\hsize
           \csname @twocolumnfalse\endcsname
\title{Mid-rapidity charge distribution in peripheral heavy ion collisions}
\author{Klaus Morawetz$^{1,2}$, Pavel Lipavsk\'y$^3$, Jacques
  Normand$^1$,
Daniel Cussol$^1$, Jean Colin$^1$   and Bernard Tamain$^1$}
\address{$^1$ LPC-ISMRA, Bld Marechal Juin, 14050 Caen, France\\
$^2$Max-Planck-Institute for the Physics of Complex Systems,
Noethnitzer Str. 38, 01187 Dresden, Germany\\
$^3$ Institute of Physics, Academy of Sciences, Cukrovarnick\'a 10,
16200 Praha 6, Czech Republic}
\maketitle
\begin{abstract}
The charge density distribution with respect to the velocity of matter
produced in peripheral heavy ion reactions around Fermi energy is
investigated. The experimental finding of enhancement of mid-rapidity
matter shows the necessity to include correlations beyond BUU which
was performed in the framework of nonlocal kinetic theory. Different
theoretical improvements are discussed. While the in--medium cross
section
changes the number of collisions, it leaves the transferred energy
almost unchanged. In contrast the nonlocal scenario changes the energy
transferred during collisions and leads to an enhancement of
mid--rapidity matter. The renormalisation of quasiparticle energies
can be included in nonlocal scenarios and leads
to a further enhancement of mid--rapidity matter distribution. This renormalisation is accompanied by a
dynamical softening of the equation of state seen in longer
oscillation periods of the excited compressional collective  mode. We
propose to include quasiparticle renormalisation by using the
Pauli--rejected collisions which circumvent the problem of backflows in
Landau theory.
Using the maximum relative velocity of
projectile and target like fragments we associate experimental events
with
impact parameters of the simulations. For peripheral collisions we find
a reasonable agreement between experiment and theory. For more central
collisions, the velocity damping is higher in one--body simulations than
observed experimentally, because of missing cluster
formations in the kinetic theory used.
\end{abstract}
\vskip2pc]

\section{Introduction}

Numerical simulations based either on the Boltzmann equation (including
the Pauli blocking it is often called the BUU equation \cite{BD88}) or
on the closely related method of quantum molecular dynamics (QMD)
\cite{SG86,A91} are extensively used to interpret experimental data from
heavy ion reactions. Due to their quasi-classical character, they offer
a transparent picture of the internal dynamics of reactions and allow
one to link observed particle spectra with individual stages of
 reactions.
Naturally, simulation results are only approximate. For example, BUU
simulations fail to describe the energy and angular distribution of
neutrons and protons in low and mid energy domain \cite{T95,Ba95,S96}.

In particular, the formation of a neck--like structure in peripheral heavy ion
reactions and the impact on the fragmentation mechanism and production of
light charged particles has been discussed for a couple of years \cite{Stuttge92,Casini93,Montoya94,ColDiT95,Lecolley95,Toke95,Stefanini95,Dempsey96,Chen96,CTGMZW98,FCD98,Larochelle99,B00}. It
has been suggested that this neck instability can be important for the
fast decomposition of matter and is probably neutron rich
\cite{SDCD97}.
Theoretical investigations
suggest that the neck is not formed in usual heavy ion  simulations starting from the Landau equation \cite{PR87,DDM90,DRS91}
or BUU equations \cite{KH94,ACC95} including additional mean field fluctuations derived
in \cite{BV85,Fl89} and tested \cite{TV85}. The inclusion of
fluctuations in the Boltzmann (BUU) equation has been investigated in
\cite{Ch95,CH96} resulting in Boltzmann-Langevin pictures
\cite{SASB90,AG90,RR90,ASBB92,ColBur93,CoCh94,ColDiT95,CTGMZW98,FCD98}.

We will take here the point of view that the fluctuations should arise
by themselves in a proper kinetic description where all relevant
correlations 
are
included in the collision integral. The collision will then cause both
a dephasing and fluctuation by itself. This procedure without
additional assumption about fluctuations has been given by the
nonlocal kinetic theory \cite{SLM96,MLSCN98,LSM99} and applied
to heavy ion  collisions in \cite{MLSK98,MT00,MTP00}. We claim
that the derived nonlocal off-set in the collision procedure induces
fluctuation in the density and consequently in the mean-field which
are similar to the one assumed ad-hoc in the approaches above. As a
first step to verify this we will investigate the formation of a neck
and the mid--rapidity emission pattern. Indeed as we will demonstrate,
the neck is much more pronounced if a nonlocal kinetic theory is used.

Recent INDRA observation shows an enhancement of emitted matter in the
region of almost zero relative velocity which means that matter is
stopped
during the reaction and stays almost at rest \cite{B00,P00}.
This enhancement of
mid--rapidity distribution can possibly be associated with a pronounced neck
formation of matter. A pioneering work on describing such mid--rapidity
emission has been done by E. Galichet and F. Gulminelli
\cite{G98}. The main problem is to find a proper selection of
experimental data \cite{L00}. The comparison between
data and simulation has been performed according to cuts in the
transverse energy. This was possible since the applied coalescence
model allows to get rid of the Fermi motion. Since we want to omit
coalescence we select the proper comparison with respect to the
maximum velocity of the projectiles with respect to the ratio of
transverse to total energy. The latter ratio gives a nearly Fermi
motion independent scaling.

We want to investigate here the peripheral heavy ion collisions and
want to discuss different theoretical improvements of the BUU
simulations. We start from the nonlocal BUU equation which includes
from  microscopic derivation the effect of binary correlations on the
collision process. By this way we obtain a nonlocal off-set of the
collision partners which account for the readjustment of the
trajectories according to the virial corrections, which would be for
hard spheres the excluded volume. This off-set induces mean-field
fluctuations similar to the above mentioned improvements of BUU. We
will give a combined picture of nonlocal off-sets and quasiparticle
renormalisation which leads to the consistent inclusion of binary
correlation on the collision integral and mean-field fluctuations.
Section II represents the formerly derived nonlocal shifts in terms of
an intuitive picture and discusses symmetries and implementation in
numerical codes. The quasiparticle renormalisation is suggested in a novel form
using the excluded events by Pauli - blocking. In section III we
present the numerical results and compare different approximations
with the experimental data. Finally in section IV we summarise and
give some outlook.

\section{Theoretical Preliminaries}\label{theory}
The need for nonlocal corrections can be stimulated by discussing the
scattering of two particles as superpositions of wave packets
\cite{SLMa96} and similar used in \cite{AH97}. The asymptotic wave packet after scattering can be written for large distance $x$ from scattering centre
\begin{eqnarray}
\phi^{\rm sc}(x,\kappa_f,t)&=&\int {d \kappa\over (2 \pi \hbar)^3} {\cal F} (x,\kappa,\kappa_f) {f(\kappa,\cos\theta)\over x} {\rm e}^{i (\kappa x -\epsilon_\kappa t)}
\label{p1}
\end{eqnarray}
with the scattering amplitude $f(\kappa,\cos\theta)$ where $\theta$ is
the angle between the relative momenta before, $\kappa$, and after the
collision, $\kappa_f$. We proceed now and expand the scattering
amplitude around the final difference momenta $\kappa_f$
\begin{eqnarray}
&&f(\kappa,\cos\theta)=|f(\kappa,\cos\theta)|{\rm e}^{i\delta(\kappa,\cos\theta)}\nonumber\\
&&=f(\kappa_f,\cos\theta) \left (1+ (\kappa-\kappa_f) \nabla_\kappa |f(\kappa,\cos\theta)|_{\kappa=\kappa_f}\right ) \nonumber\\&&
\times{\rm e}^{i (\kappa-\kappa_f) \nabla_\kappa \delta(\kappa,\cos\theta)_{\kappa=\kappa_f}}.\nonumber\\
&&
\end{eqnarray}
The derivative of the phase $\delta$ leads now to the definition of the effective space shifts $\Delta^f$ and the time shift $\Delta_{||}$
\begin{eqnarray}
2\Delta^f&=&\nabla_\kappa\delta|_{\kappa=\kappa_f}={\kappa_f\over |\kappa_f|}\partial_{\kappa_f} \delta+{{\kappa\over |\kappa|}-\cos\theta{\kappa_f\over |\kappa_f|}
\over |\kappa_f|}\partial_{\cos\theta}\delta
\nonumber\\
&\equiv& {\kappa\over m}\Delta_{\|}+({\kappa\over |\kappa|}-\cos\theta{\kappa_f\over |\kappa_f|})\Delta_{\perp}
\label{pp}
\end{eqnarray}
where we denoted the shifts corresponding to the direction of $\kappa$
as $\|$ and $\perp$. Rewriting (\ref{p1}) we obtain
\begin{eqnarray}
\phi^{\rm sc}(x,\kappa_f,t)&=&{f(\kappa_f,\cos\theta)\over x}\nonumber\\
&\times &\int {d\kappa\over (2 \pi \hbar)^3} \tilde
{\cal F}(x,\kappa,\kappa_f) {\rm e}^{i\kappa \cdot (x+2 \Delta^f)}{\rm
  e}^{-i \epsilon_{\kappa} (\Delta_{\|}+t)}.
\nonumber\\&&
\end{eqnarray}
We observe three effects of scattering on the asymptotics: (i) a genuine time delay $\Delta_{\|}$, (ii) an effective displacement of the two colliding particle of $\Delta^f$ with respect to the centre of mass and (iii) a modification of scattering probability
$\tilde{\cal F}={\cal F} (1+ (\kappa-\kappa_f) \nabla_\kappa |f(p,\cos\theta)|_{\kappa=\kappa_f})$.

The effect of non-local collisions on the dynamics of heavy ion reactions
has been studied already within a cascade model \cite{H81}. For a simple
hard-sphere approximation of nucleon-nucleon collisions, Halbert has
demonstrated that density patterns of $^{20}$Ne$+^{238}$U reactions are
sensitive to local or non-local treatment of collisions. Malfliet
\cite{M83} also found disturbing that all dynamical models rely more or
less on the use of the local approximation of binary collisions, because
the local approximation neglects a contribution of the collision flux to
both material relations which control the hydrodynamic motion during the
reaction, the compressibility and the share viscosity. To include the
collision flux, Malfliet incorporated hard-sphere non-local collisions
into the BUU simulation code. Recently, this approximation has been used
by Kortemeyer, Daffin and Bauer \cite{KDB96}.

The hard-sphere approximation of a non-local collision is sufficient for
the above mentioned discussions of trends, however, it cannot be used in
realistic studies. This ad hoc approximation has been used not only for
its simplicity but also because of lack of a first principle theory
offering quantum mechanical displacements which would generalise the
classical hard-sphere displacements proposed by Enskog. As far as we
know, till recently there was no non-local theory of binary collisions
devoted to the nuclear matter. In literature, there are closely related
quantum theories of binary collisions developed for moderately dense
gases \cite{NTL91,H90}, these, however, treat non-local collisions via
gradient contributions to the scattering integral. The gradient form is
suitable for hydrodynamic expansions studied in the chemical physics,
but is very inconvenient for numerical simulation and thus have never
been employed for heavy ion reactions.

Recent theoretical studies have filled this gap in theory. Danielewicz
and Pratt \cite{DP96} pointed out that the collision delay can be used
as a convenient tool to describe the virial corrections to the equation
of state for the gas of quasiparticles. Although their discussion is
limited to the equilibrium, it marks a way how to introduce virial
corrections also to dynamical processes. The kinetic equation for
quasiparticles with non-instantaneous and non-local scattering integral
has been derived in \cite{SLM96,LSM97} as a systematic quasi-classical limit
of non-equilibrium Green's functions in the Galitskii--Feynman
approximation. It has been shown that the gradient
corrections to the scattering integral can be rearranged into a form
of a collision delay and space displacements reminiscent of classical hard
spheres, i.e., into a form suitable for numerical simulations.

In this contribution we will put these ideas of nonlocalities
on the firm ground using the quantum kinetic equation with
nonlocal scattering integrals which was derived from quantum
statistics \cite{SLM96,LSM97} to show how the effect of nonlocalities play
a role in simulations of heavy ion reactions and compare them with experiment.

\subsection{Nonlocal kinetic theory}
\begin{figure}
  \psfig{figure=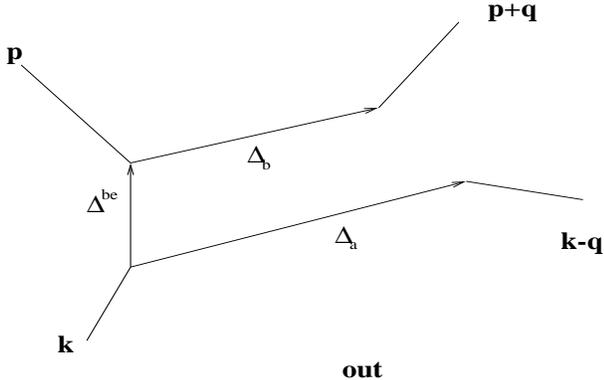,width=8cm,height=5cm}
\caption{A nonlocal binary collision according to Eq. (\protect\ref{delta}).
\label{soft}}
\end{figure}

The scattering integral of the non-local kinetic equation derived in
\cite{SLM96} corresponds to a following picture of a collision as seen
in Figure~\ref{soft}.

Assume
that two particles, $a$ and $b$, of initial momenta $k$ and $p$ start to
collide at time instant $t$ being at coordinates $r_a$ and $r_b$. Due to
a finite range of the interaction, at the beginning of collision
particles are displaced by $r_b-r_a=\Delta^{\rm be}$. The collision has
a finite duration $\Delta_t$, i.e., it ends at $t+\Delta_t$. During the
collision, both particles move so that their end coordinates differ from
those at the beginning, $r_a'-r_a=\Delta_a$ and $r_b'-r_b=\Delta_b$. The
particle $a$ transfers a momentum $q$ to the particle $b$, therefore
their relative momentum changes from $\kappa=\frac 1 2 (k-p)$ to
$\kappa'=\frac 1 2 (k-p)-q$. Their sum momentum is modified by an
external field acting on the colliding particles during the collision
going from $K=k+p$ to $K'=k+p+\Delta_K$. The same field changes the sum
energy of colliding particles from $E=\epsilon_a+\epsilon_b$ to $E'=
\epsilon_a'+\epsilon_b'=\epsilon_a+\epsilon_b+\Delta_E$.

The values of $\Delta$'s are given by derivatives of the scattering
phase shift \mbox{$\phi={\rm Im\ ln}T_R(\Omega,k,p,q,t,r)$},
\begin{eqnarray}
\Delta_t&=&\left.{\partial\phi\over\partial\Omega}\right|_E,
\nonumber\\
\Delta^{\rm be}&=&\left({\partial\phi\over\partial p}-
{\partial\phi\over\partial q}-{\partial\phi\over\partial k}\right)_E,
\nonumber\\
\Delta_a&=&\left.-{\partial\phi\over\partial k}\right|_E,
\nonumber\\
\Delta_b&=&\left.-{\partial\phi\over\partial p}\right|_E,
\nonumber\\
\Delta_K&=&\left.{\partial\phi\over\partial r}\right|_E,
\nonumber\\
\Delta_E&=&\left.-{\partial\phi\over\partial t}\right|_E.
\label{delta}
\end{eqnarray}
Note that energy $\Omega$ enters as an independent quantity so that one
needs to know the scattering phase shift out of the energy shell. The
on-shell energy, $\Omega=E$, is substituted after derivatives are taken.
Since experiments provide us only with the on-shell values of the
scattering phase shift $\phi$, set (\ref{delta}) of $\Delta$'s cannot be
derived directly from experimental values but only indirectly via
off-shell T-matrix constructed from the interaction potential. Below we
show how to circumvent this complication.

\subsection{Instantaneous approximation}
It is our intention to incorporate these features of collisions into the
BUU (or QMD) simulation codes. The selfconsistent evaluation of all
$\Delta$'s for all collisions would be too demanding. We employ two
kinds of additional approximations. First
we neglect the medium effect on binary
collisions, i.e., use the well known free-space T-matrix. Second, we
rearrange the scattering integral into an instantaneous but non-local
form. This instantaneous form parallels hard-sphere-like collisions what
allow us to employ computational methods developed within the theory of
gases \cite{AGA95} in a similar way to \cite{KDB96}.

In the instantaneous approximation we let particles to make a sudden
jump from their in-coming trajectories into their out-going ones so that
at time $t+\Delta_t$ particles arrive at the correct coordinates, $r_a'$
and $r_b'$, with the correct momenta, $\kappa'$ and $K'$. Accordingly,
in the asymptotic region, after $t+\Delta_t$, there is no distinction
between the non-instantaneous and instantaneous pictures.

The actual time instant at which the jump happens does not influence the
asymptotic states, however, two particular choices of the time instant
are important with respect to the implementation into simulations. The
first one is the time at which the simulation code selects two particles
as adepts for the collision usually at the point of closest
approach.
We will discuss this time later when we
specify the simulation scheme. The second one is the central time,
$\tilde t=t+{1\over 2}\Delta_t$, for which the instantaneous
approximation maintains the space and time symmetries of the
non-instantaneous collision. As we will see, these symmetries allow one
to derive the value of the sudden jump from experimental phase shifts.

The condition that the sudden jump correctly mimics the
non-instantaneous process is naturally met if one extrapolates the
in-coming and out-going trajectories from known coordinates and momenta
at $t$ and $t+\Delta_t$, respectively, to the central time $\tilde t$.
Doing so one finds that extrapolated coordinates just before and after
the sudden jump read
\begin{eqnarray}
\tilde r_a&=&r_a+{k\over 2m}\Delta_t,
\nonumber\\
\tilde r_b&=&r_b+{p\over 2m}\Delta_t,
\nonumber\\
\tilde r_a'&=&r_a'-{k-q\over 2m}\Delta_t=
r_a+\Delta_a-{k-q\over 2m}\Delta_t,
\nonumber\\
\tilde r_b'&=&r_b'-{p+q\over 2m}\Delta_t=
r_b+\Delta_b-{p+q\over 2m}\Delta_t.
\label{da}
\end{eqnarray}
In extrapolations of momenta we neglect Coulomb forces so that we assume
that protons and neutrons are driven by the same force from the Skyrme
potential $U$, $F=-{\partial U\over\partial r}$. In this case, the
relative momentum remains unchanged and the force affects only the sum
momentum,
\begin{eqnarray}
\tilde\kappa&=&\kappa,
\nonumber\\
\tilde K&=&K+F\Delta_t,
\nonumber\\
\tilde\kappa'&=&\kappa'=\kappa-q,
\nonumber\\
\tilde K'&=&K'-F\Delta_t=K+\Delta_K-F\Delta_t,
\label{db}
\end{eqnarray}
Finally, energies at the extrapolated phase-space points are
\begin{eqnarray}
\tilde\epsilon_a&=&{\tilde k^2\over 2m}+
U\left(\tilde r_a,t+{\Delta_t\over 2}\right)=
\epsilon_a+{\partial U\over\partial t}{\Delta_t\over 2},
\nonumber\\
\tilde\epsilon_b&=&{\tilde p^2\over 2m}+
U\left(\tilde r_b,t+{\Delta_t\over 2}\right)=
\epsilon_b+{\partial U\over\partial t}{\Delta_t\over 2},
\nonumber\\
\tilde\epsilon_a'&=&{(\tilde k-q)^2\over 2m}+
U\left(\tilde r_a',t+{\Delta_t\over 2}\right)=
\epsilon_a'-{\partial U\over\partial t}{\Delta_t\over 2},
\nonumber\\
\tilde\epsilon_b'&=&{(\tilde p+q)^2\over 2m}+
U\left(\tilde r_b',t+{\Delta_t\over 2}\right)=
\epsilon_b'-{\partial U\over\partial t}{\Delta_t\over 2}.
\label{dc}
\end{eqnarray}
In rearrangement we have used that the increase of the kinetic energy,
say ${\tilde k^2\over 2m}-{k^2\over 2m}={1\over 2m}\left(k+{1\over 2}F
\Delta_t\right)^2-{k^2\over 2m}={kF\over 2m}\Delta_t$, is compensated
by the decrease of the potential energy along the trajectory, $U(\tilde
r_a,t)-U(r_a,t)={\partial U\over\partial r}{k\over 2m}\Delta_t=-F{k\over
2m}\Delta_t$. In a stationary potential, the compensation of kinetic and
potential energies reflects the energy conservation. The second order
corrections in $\Delta$'s are neglected.

Using the extrapolated quantities, we can define a new set of effective
$\Delta$'s corresponding to the instantaneous picture,
\begin{eqnarray}
\tilde\Delta^{\rm be}&=&\tilde r_b-\tilde r_a=\Delta^{\rm be}-
\left({k\over m}-{p\over m}\right){\Delta_t\over 2},
\nonumber\\
\tilde\Delta_a&=&\tilde r_a'-\tilde r_a=\Delta_a-
\left({k-q\over m}+{k\over m}\right){\Delta_t\over 2},
\nonumber\\
\tilde\Delta_b&=&\tilde r_b'-\tilde r_b=\Delta_b-
\left({p+q\over m}+{p\over m}\right){\Delta_t\over 2},
\label{tildelta}
\end{eqnarray}

The space displacements $\tilde \Delta^{\rm be}$, $\tilde \Delta_a$
and $\tilde \Delta_b$ can
be expressed in terms of the on-shell scattering phase shift defined as
\begin{equation}
\tilde\phi(k,p,q,r,t)=\left.\phi\right|_{\Omega={1\over 2}(\epsilon_a+
\epsilon_b+\epsilon_a'+\epsilon_b')}.
\label{tilphi}
\end{equation}
From (\ref{delta}) and (\ref{tildelta}) one can directly check that
\begin{eqnarray}
\tilde\Delta^{\rm be}&=&{\partial\tilde\phi\over\partial p}-
{\partial\tilde\phi\over\partial q}-{\partial\tilde\phi\over\partial k},
\nonumber\\
\tilde\Delta_a&=&-{\partial\tilde\phi\over\partial k},
\nonumber\\
\tilde\Delta_b&=&-{\partial\tilde\phi\over\partial p}.
\label{redelta}
\end{eqnarray}
Effective displacements (\ref{redelta}) can be evaluated from the
experimentally observed scattering phase shifts. Before we turn to this
pragmatic question, it is profitable to enlighten the conservation laws
and symmetries of the collision processes.

\subsection{Conservation laws}
The extrapolated momentum and energy gains vanish,
\begin{eqnarray}
\tilde\Delta_K&=&\tilde K'-\tilde K=\Delta_K-2F\Delta_t=0,
\nonumber\\
\tilde\Delta_E&=&\tilde\epsilon_a'+\tilde\epsilon_b'-\tilde\epsilon_a-
\tilde\epsilon_b=\Delta_E-2{\partial U\over\partial t}\Delta_t=0.
\label{tildelta0}
\end{eqnarray}
To show this we use that for a collision of isolated nucleons, the
scattering phase shift depends only on the initial and final momenta,
$\kappa$ and $\kappa'$, while sum momentum $K$ and the Skyrme potential
only shifts the energy bottom,
\begin{equation}
\phi(\Omega,k,p,q,r,t)=\phi\left(\Omega-{(k+p)^2\over 4m}-2U,k-p,q
\right).
\label{phi}
\end{equation}
The time derivative which results in the energy gain $\Delta_E$ thus can
be expressed via the energy derivative and from (\ref{delta}) one finds
that $\Delta_E=2{\partial U\over\partial t}\Delta_t$, therefore $\tilde
\Delta_E=0$. The space dependency also enters $\phi$ only via the energy
argument, therefore $\Delta_K=2F\Delta_t$ or $\tilde\Delta_K=0$. The
cancellation of both shifts simplifies the energy and momentum
conservation to its form commonly used for instantaneous collisions.
Briefly, in the instant collision the Skyrme potential has no time to
pass any energy and momentum to colliding nucleons.

An additional simplification follows from the continuity of the centre of
mass motion. This requires $\tilde r_a'+\tilde r_b'=\tilde r_a+\tilde
r_b$ or $\tilde\Delta_a+\tilde\Delta_b=0$. This relation is satisfies by
displacements (\ref{redelta}). Indeed, in the approximation of isolated
collision (\ref{phi}), the on-shell energy argument reduces,
\begin{eqnarray}
{1\over 2}(\epsilon_a+\epsilon_b+\epsilon_a'+\epsilon_b')&-&
{(k+p)^2\over 4m}-2U
\nonumber\\
&=&{(k-p)^2\over 8m}+{(k-p-2q)^2\over 8m}.
\label{redarg}
\end{eqnarray}
The on-shell scattering phase shift (\ref{tilphi}) then does not depends
on the sum momentum $k+p$, therefore the derivatives with respect to $k$
and $p$ are mutually connected,
\begin{equation}
{\partial\tilde\phi\over\partial k}=-
{\partial\tilde\phi\over\partial p}.
\label{interl}
\end{equation}
According to (\ref{redelta}), the displacement of the particle $b$ is
opposite to the displacement of the particle $a$,
\begin{equation}
\tilde\Delta_b=-\tilde\Delta_a.
\label{interlsym}
\end{equation}

\subsection{Rotational symmetry}
The symmetries are best seen in the barycentric representation in which
the phase shift is a function of the initial and final relative momenta,
$\tilde\phi(k-p,q)\equiv\tilde\phi(\kappa,\kappa')$, where $\kappa=
{1\over 2}(k-p)$ and $\kappa'={1\over 2}(k-p-2q)$. From substitution
into the barycentric framework one obtains
\begin{eqnarray}
{\partial\over\partial k}&=&{\partial\kappa\over\partial k}
{\partial\over\partial\kappa}+{\partial\kappa'\over\partial k}
{\partial\over\partial\kappa'}={1\over 2}{\partial\over\partial\kappa}+
{1\over 2}{\partial\over\partial\kappa'}
\nonumber\\
{\partial\over\partial p}&=&{\partial\kappa\over\partial p}
{\partial\over\partial\kappa}+{\partial\kappa'\over\partial p}
{\partial\over\partial\kappa'}=-{1\over 2}{\partial\over\partial\kappa}-
{1\over 2}{\partial\over\partial\kappa'}
\nonumber\\
{\partial\over\partial q}&=&{\partial\kappa\over\partial q}
{\partial\over\partial\kappa}+{\partial\kappa'\over\partial q}
{\partial\over\partial\kappa'}=-{\partial\over\partial\kappa'},
\label{parder}
\end{eqnarray}
therefore the displacements in terms of relative momenta read
\begin{eqnarray}
\tilde\Delta_a&=&-{1\over 2}{\partial\tilde\phi\over\partial\kappa}-
{1\over 2}{\partial\tilde\phi\over\partial\kappa'},
\nonumber\\
\tilde\Delta^{\rm HS}&=&\tilde\Delta_b+\tilde\Delta^{\rm be}=
\tilde\Delta^{\rm be}-\tilde\Delta_a
\nonumber\\
&=&-{1\over 2}{\partial\tilde\phi\over\partial\kappa}+
{1\over 2}{\partial\tilde\phi\over\partial\kappa'}.
\label{bardelta}
\end{eqnarray}
Apparently, the $\tilde\Delta_a$ does not change under replacement of
the initial and final momenta, $\kappa\longleftrightarrow\kappa'$. As a
complementary displacement we have introduced $\tilde\Delta^{\rm HS}$
instead of
$\tilde \Delta^{\rm be}$ which reverses its orientation under the
replacement of initial and final momenta.

For central forces, the scattering phase shift has to satisfy the
rotational symmetry, therefore it depends only on the deflection angle
$\theta$
\begin{equation}
\cos\theta={\kappa\kappa'\over|\kappa||\kappa'|}
\label{theta}
\end{equation}
and amplitudes of initial and final relative momenta,
\begin{equation}
\tilde\phi(\kappa,\kappa')=\tilde\phi(\cos\theta,|\kappa|,|\kappa'|).
\label{rotphi}
\end{equation}
The vector derivatives follow from ${\partial\over\partial\kappa}
\cos\theta=\kappa'{1\over|\kappa||\kappa'|}-\kappa{\cos\theta\over|
\kappa|^2}$ and ${\partial\over\partial\kappa}|\kappa|={\kappa\over|
\kappa|}$ as
\begin{equation}
{\partial\tilde\phi\over\partial\kappa}=\kappa\left({1\over|\kappa|}
{\partial\tilde\phi\over\partial|\kappa|}-
{\cos\theta\over|\kappa||\kappa'|}{\partial\tilde\phi\over\partial
\cos\theta}\right)+\kappa'{1\over|\kappa||\kappa'|}
{\partial\tilde\phi\over\partial\cos\theta}.
\label{dersim}
\end{equation}
The derivative with respect to the final momentum $\kappa'$ is obtained
from (\ref{dersim}) via the interchange $\kappa\longleftrightarrow
\kappa'$.

\subsection{Time-reversal symmetry}
As mentioned above, we want to express the effective displacements in
terms of observable scattering phase shifts. This will make possible to
circumvent the uncertainty about interaction potentials and to supply
the simulation codes directly with experimental values.

Although the phase shift $\tilde\phi$ does not depend on the energy
$\Omega$, it is still not experimentally known for general momentum $q$
but only for momenta which satisfy the energy conservation, $(k-p)^2=
(k-p-2q)^2$. Due to time-reversal symmetry of the collision process,
it is possible to express the effective displacements in terms of
derivatives along this momentum shell.

The experimentally available values of the scattering phase shift are
restricted to the shell $|\kappa'|=|\kappa|$,
\begin{equation}
\phi_{\rm exp}(\cos\theta,|\kappa|)=\left.\tilde\phi(\cos\theta,
|\kappa|,|\kappa'|)\right|_{|\kappa'|=|\kappa|}.
\label{phiexp}
\end{equation}
In (\ref{dersim}) we need separate derivations with respect to
$|\kappa|$ and $|\kappa'|$. Fortunately, due to the time and space
reversal symmetries, the scattering phase shift is a symmetric function
of $|\kappa|$ and $|\kappa'|$,
\begin{equation}
\tilde\phi(\cos\theta,|\kappa|,|\kappa'|)=
\tilde\phi(\cos\theta,|\kappa'|,|\kappa|).
\label{phirev}
\end{equation}
Since we need to find the derivatives only for $|\kappa'|=|\kappa|$ and
symmetry (\ref{phirev}) implies
\begin{equation}
\left.{\partial\over\partial|\kappa'|}\tilde\phi(|\kappa|,|\kappa'|)
\right|_{|\kappa'|=|\kappa|}=
\left.{\partial\over\partial|\kappa|}\tilde\phi(|\kappa|,|\kappa'|)
\right|_{|\kappa'|=|\kappa|}
\label{phirevder}
\end{equation}
we obtain a direct link to observable scattering phase shift
\begin{equation}
\left.{\partial\over\partial|\kappa'|}\tilde\phi(\cos\theta,|\kappa|,
|\kappa'|)\right|_{|\kappa'|=|\kappa|}={1\over 2}
{\partial\over\partial|\kappa|}\phi_{\rm exp}(\cos\theta,|\kappa|).
\label{phiexpder}
\end{equation}

\subsection{Displacements in simulations}
Now we are ready to evaluate the effective displacements from
experimental scattering phase shifts. From (\ref{bardelta}),
(\ref{dersim}) and (\ref{phiexpder}) we find
\begin{eqnarray}
\tilde\Delta_a&=&-{\kappa+\kappa'\over 2|\kappa|^2}
\left({|\kappa|\over 2}{\partial\phi_{\rm exp}\over\partial|\kappa|}-
\left(\cos\theta-1\right){\partial\phi_{\rm exp}\over\partial\cos\theta}
\right),
\nonumber\\
\tilde\Delta^{\rm HS}&=&-{\kappa-\kappa'\over 2|\kappa|^2}
\left({|\kappa|\over 2}{\partial\phi_{\rm exp}\over\partial|\kappa|}-
\left(\cos\theta+1\right){\partial\phi_{\rm exp}\over\partial\cos\theta}
\right).
\label{deltaf}
\end{eqnarray}
In agreement with their even/odd symmetry under interchange of initial
and final momenta, the $\tilde\Delta_a$ is proportional to the sum
$\kappa+\kappa'$ while $\tilde\Delta^{\rm HS}$ to the difference
$\kappa+\kappa'$. From the energy conservation, $(\kappa+\kappa')
(\kappa-\kappa')=\kappa^2-\kappa'^2=0$, follows that these two vectors
are orthogonal. Before we present actual values of displacements, it is
profitable to compare their form with the model of hard spheres and the
approximation by the collision delay.

For classical hard spheres of radius $R$, the scattering phase shift
depends only on the transferred momentum $q=\kappa-\kappa'$ as,
$\phi^{\rm HS}=\pi-2|q|R$. From (\ref{deltaf}) than follows that
$\tilde\Delta_a=0$ and $\tilde\Delta^{\rm HS}={q\over|q|}2R$. The
hard-sphere approximation used in model studies \cite{H81,M83,KDB96}
thus neglects $\tilde\Delta_a$ and uses a constant approximation of the
amplitude of the other displacement, $\left|\tilde\Delta^{\rm HS}\right|
=2R$. This amplitude is conveniently evaluated from the cross section,
$\sigma=\pi R^2$.

\begin{figure}
\begin{minipage}{7cm}
\parbox[t]{7cm}{
  \psfig{figure=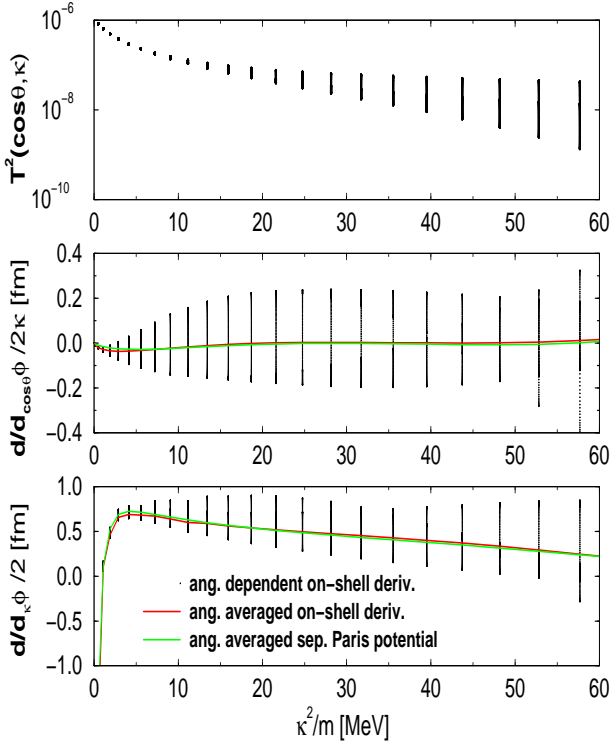,width=10cm,height=8cm,angle=-90}}
\end{minipage}
\caption{The effective displacement as a function of the deflection
angle and the kinetic energy, $\kappa^2/m$, in the barycentric
coordinate system. The columns of dots show the spread of components
with deflection angle. The lines show the angle-averaged values. The
amplitude of the T-matrix is presented in the top section to indicate
the weight of individual processes. The orthogonal component,
$\overline{\Delta_\perp}/2\equiv d \phi/d{\cos\theta}/2\kappa$ shown in the
middle section, has appreciably smaller values than the parallel
component, $\overline{\Delta_\|}/2m\equiv d \phi/d \kappa/2$ shown in
the bottom section.}
\label{ixf1}
\end{figure}

The approximation discussed by Danielewicz and Pratt \cite{DP96} deals
only with the collision delay defined according to Wigner as the energy
derivative of the scattering phase shift. This approximation is obtained
from (\ref{deltaf}) if one neglects the derivative with respect to the
deflection angle, therefore $\tilde\Delta_a=-{\kappa+\kappa'\over 2m}
{\partial\phi_{\rm exp}\over\partial E}$ and $\tilde\Delta^{\rm HS}=-
{\kappa-\kappa'\over 2m}{\partial\phi_{\rm exp}\over\partial E}$, where
$E=|\kappa|^2/m$ is energy in the barycentric system.

Numerical values of these two contributions are compared in
Fig.~\ref{ixf1}. The dots in the vertical line show a spread of values
due to the angular dependence, the curves show values averaged over
deflection angles with the weight given by the differential cross
section displayed in the top section. The parallel component, shown in
the bottom section, has a typical value of 0.5~fm. The negative large
values below 3~MeV can be ignored since corresponding processes have
very small rates due to the Pauli blocking. The perpendicular component,
shown in the middle section, has about three-times smaller values,
moreover it tends to average out.
For energies above $10$MeV,
the displacements can be well approximated by a constant value, as it is
case of the hard-sphere model. Moreover, the amplitude of the
displacement is close to the estimate based on the differential cross
section, in spate conceptual difference between both concepts. Our
results thus confirm that estimates used in \cite{H81,M83,KDB96,DP96} are
quite reasonable.

\begin{figure}
  \psfig{figure=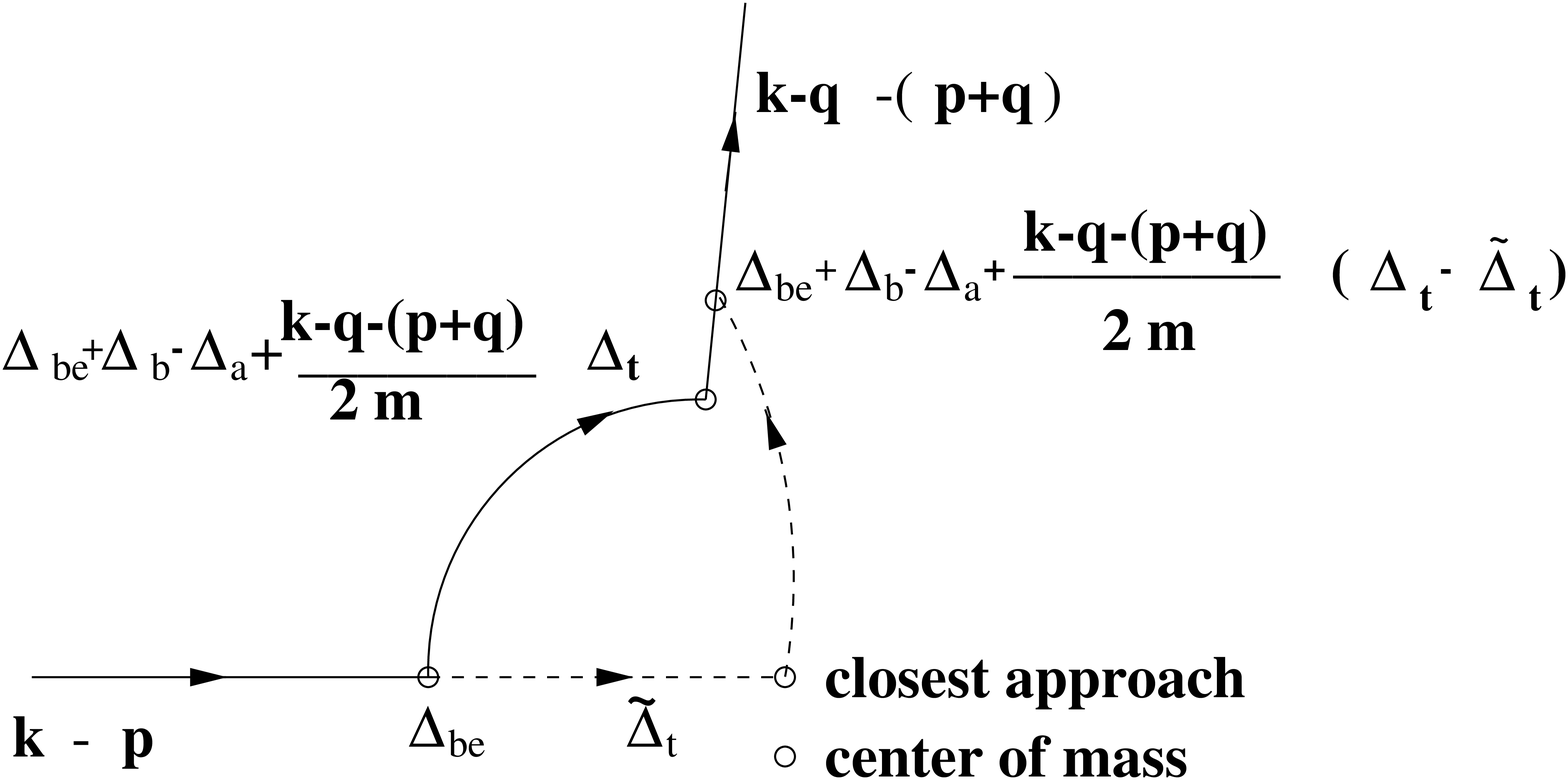,width=8cm,height=5cm}
\vspace{0.5cm}
\caption{A nonlocal binary collision (solid line) together with the scenario of sudden jump at the closest approach.
\label{softc}}
\end{figure}

In \cite{MLSCN98} we have used as the time of instant jump the time of
closest approach.
This distance is different from the distance $\Delta^{\rm be}$
required from the equivalent scattering scenario presented in figure
\ref{softc} as solid line.We consider now the time required to travel
from $\Delta_{\rm be}$ to the distance of closest approach
$\tilde\Delta_t={m\over 2 \kappa^2} \kappa \Delta_{\rm be}$ in analogy
to \cite{T81}. Within this scenario we are allowed to jump at the
point of closest approach to the final asymptotics (\ref{da}) and
(\ref{db})with the additional distance the particle travel during
$\tilde\Delta_t$. The effective final jump in the barycentric frame is
\be
\Delta^f&=&\frac 1 2 (\Delta_a-\Delta_b-\Delta^{\rm be})-{\kappa'\over
  m}(\Delta_t-\tilde\Delta_t)\nonumber\\
&=&\frac 1 2 (\tilde\Delta_a-\tilde\Delta^{\rm HS})+{\kappa'\over
  2 |\kappa|^2} \kappa \cdot (\tilde\Delta_a+\tilde\Delta^{\rm HS})
\label{dd}
\ee
where the same compensation of off-shell derivatives by $\Delta_t$
occurs as described before when jumping at the centre time $\tilde t$.
Please remark that (\ref{dd}) agrees with (\ref{pp}).

\subsection{Renormalisation of quasiparticle energies}\label{theoryren}

So far we have discussed the nonlocal shifts as if there were free
classical particles. The interaction affects, however, the free motion
of particles between individual collisions. The dominant effect is due 
to mean-field forces which bind the nucleus together, accelerating
particles close to the surface towards the centre. These forces are 
conveniently included via potentials of Skyrme and Hartree type.

Beside forces, the interaction also modifies the velocity with which a
particle of a given momentum propagates in the system. This effect is 
known as the mass renormalisation. A numerical implementation of the
renormalised mass is rather involved since a plain use of the
renormalised mass instead of the free one leads to incorrect currents.
Within the Landau concept of quasiparticles, this problem is cured by 
the back flow, but it is not obvious how to implement the back flow 
within the BUU simulation scheme. In our studies, we circumvent the 
problem of back flows using explicit zero-angle collisions to which we 
add a non-local correction.

\subsubsection{One-dimensional system of fixed scatterers}
A link between the renormalisation of the mass and zero-angle collision
has been already pointed out by Landau. We find it instructive to
describe this mechanism first for a simple one-dimensional system
of randomly distributed barriers. Tunnelling then corresponds to
the zero-angle scattering and reflection to a dissipative
collision. We focus on the piece of quasi-free trajectory, i.e., the
trajectory between two successive dissipative collisions.

A tunnelling through barriers speeds up or slow down the mean velocity
of particles. To simulate this effect on the motion, at each tunnelling
we shift a particle by a displacement $D$ in the direction of its 
motion. For simplicity we take the amplitude of $D$ as constant. After
time $t$ the particle moves over a distance
\begin{equation}
x = t{k\over m}+N\,D,
\label{mr1}
\end{equation}
where $k/m$ is its velocity between tunnellings and $N$ is the number
of barriers on the trajectory of length $x$. In average $N=cx$, where
$c$ is concentration of barriers. The mean velocity of a particle then
reads
\begin{equation}
v={x\over t}={k\over m}+cD{x\over t}={k\over m(1-cD)}
\approx {k\over m}(1+cD).
\label{mr2}
\end{equation}
For the three-dimensional system with particle-particle interactions, 
the corrections to the mean velocity will be limited to the linear 
approximation.

Relation (\ref{mr2}) can be compared with the velocity of a particle
evaluated within the renormalised mass $m^*$,
\begin{equation}
v={k\over m^*},
\label{mr3}
\end{equation}
from which follows
\begin{equation}
{m\over m^*}=1+cD.
\label{mr4}
\end{equation}
We can use (\ref{mr4}) to fit $D$ so that the known renormalised mass
$m^*$ is reproduced. 

\subsubsection{Three-dimensional Fermi liquid}
In the classical three-dimensional system all collisions have a finite
deflection angle. In the quantum system, however, there are zero-angle
collisions which represent an interference between scattering states
and the incoming state of the interaction. In the dialect of the
perturbative expansion one can say that the particle makes a detour
from its trajectory in the phase space but nowhere on the detour it
reaches the energy shell and thus it has to return back. The detour
causes a delay expressed by the shift $D$. In the Fermi liquid, the
Pauli exclusion principle blocks a majority of phase space cell on
the energy shell so that the zero-angle collisions dominate over the
dissipative events. We will use these blocked events to simulate the
renormalisation of the mass.

Unlike in the simple one-dimensional scattering on fixed defects, the
displacement $D$ is a vector oriented along the difference momentum,
\begin{equation}
D=|D|{k-p\over|k-p|},
\label{mrF1}
\end{equation}
where $k$ is a momentum of the assumed particle while $p$ belongs to
its partner in the prohibited collision. In general, the displacement
$|D|$ is a function of $k$ and $p$. For simplicity we assume this
function as a constant and fit its value to the mass renormalisation
at the Fermi surface.

For the fitting of the displacement $D$ we assume the zero temperature
at which all real collisions are blocked by the Pauli exclusion
principle so that all binary encounters contribute to the
renormalisation. The mean velocity of the particle is then given by its
free motion and the mean value of the displacements per the time unit,
\begin{equation}
v={k\over m}+\sum_a\int{dp\over(2\pi\hbar)^3}\sigma{|k-p|\over m}
f_p|D|{k-p\over|k-p|}.
\label{mrF2}
\end{equation}
The mean value of displacements is proportional to the frequency of
binary entertainments, i.e., it is the sum of integrals over 
distributions of protons and neutrons weighted with the scattering 
cross section $\sigma$ and their relative velocity to the observed 
particle.

With a good approximation the cross section $\sigma$ is independent of
energy so that one can easily evaluate the integral in (\ref{mrF2}),
\begin{equation}
v={k\over m}+n\sigma|D|{k\over m}-n\sigma|D|\langle v\rangle.
\label{mrF2ev}
\end{equation}
The renormalised velocity thus depends on the density,
\begin{equation}
n=\sum_a\int{dp\over(2\pi\hbar)^3}f_p={2p_F^3\over 3\pi^2\hbar^3},
\label{mrF3}
\end{equation}
and the mean velocity of the nuclear matter, 
\begin{equation}
\langle v\rangle=
{1\over n}\sum_a\int{dp\over(2\pi\hbar)^3}f_p{p\over m}.
\label{mrF10}
\end{equation}

For the system in rest, $\langle v\rangle=0$, we find the mass
renormalisation,
\begin{equation}
{m\over m^*}=1+n\sigma |D|.
\label{mrF7}
\end{equation}
One can see that this formula has the same interpretation as the
one-dimensional case (\ref{mr4}), because $n\sigma$ is the average
number of scatterers on the trajectory of unitary length. 

Formula (\ref{mrF7}) allows us to fit $|D|$ from known value of the 
effective mass. For $m^*:m=3:4$, $\sigma=40$~mb and $n=0.16$~fm$^{-3}$ 
one finds value $|D|=0.5$~fm. This value is very close to the nonlocal
correction in dissipative collisions, see Fig.~\ref{ixf1}.

The quasiparticle velocity (\ref{mrF2ev}) relates to the quasiparticle
energy by $v=\partial \epsilon_k/\partial k$.
For the moving nuclear matter $\langle v\rangle\ne 0$ one finds
\begin{equation}
\epsilon_k={k^2\over 2m}+n\sigma |D|{(k-m\langle v\rangle)^2\over 2m}.
\label{mrF11}
\end{equation}
An approximation of this structure is commonly used in simple 
applications of the Landau concept of quasiparticles.

\subsection{Summary an simulation schema}

The derived effective nonlocal collision procedure is easily
incorporated in the usual collision simulation by an additional
advection step. The quasiparticle renormalisation and effective mass
is found to be possible to incorporate by the same advection step but
performed for the events rejected normally by Pauli-blocking.

Finally, we would like to comment on properties of the proposed
simulation scheme. The renormalisation depends on the distribution of
particles in surrounding medium. It has four nice properties: (i) the
renormalisation vanishes as the local density goes to zero, (ii) the
renormalisation vanishes when a high temperature closes the Luttinger
gap because all collisions will be at finite angles, (iii) the
anisotropy of the quasiparticle velocity in a presence of a non-zero
current in medium is automatically covered, and (iv) the backflows
connected to the mass renormalisation are covered because both
particles jump keeping the centre of mass fixed. Last but not least,
the simulation does not require to introduce new time-demanding
procedures, one can simply use the scattering events which are merely
rejected in standard simulation codes by Pauli-blocking.

\section{Numerical results}

Let us discuss the proposed correction to the local and ideal (no
quasiparticle renormalisation) Boltzmann (BUU) simulation. First we
introduce the pure nonlocal corrections and then we discuss the
quasiparticle renormalisation.

\begin{figure}
\psfig{file=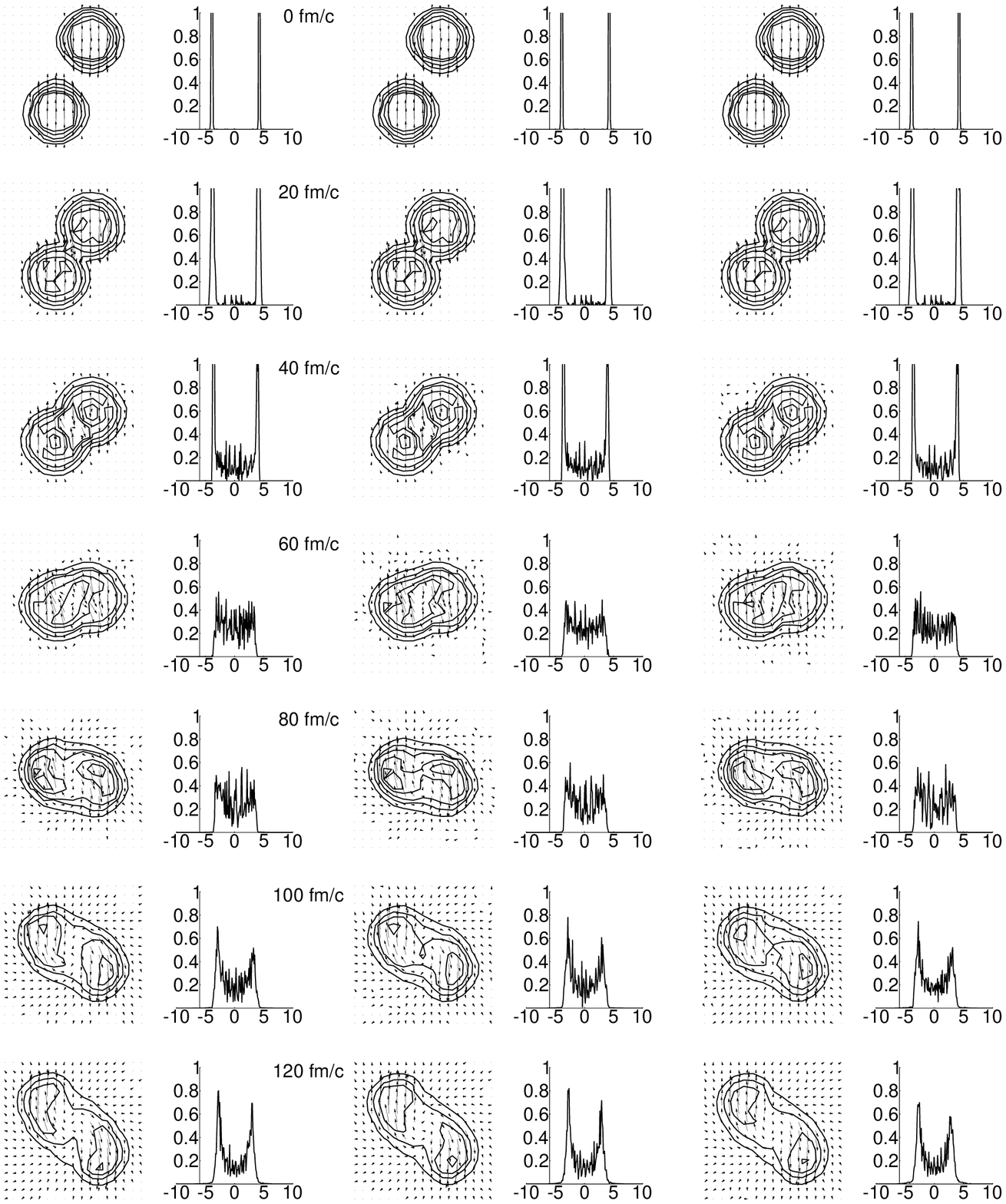,height=17cm,angle=0}
\vspace{0.2cm}

\psfig{file=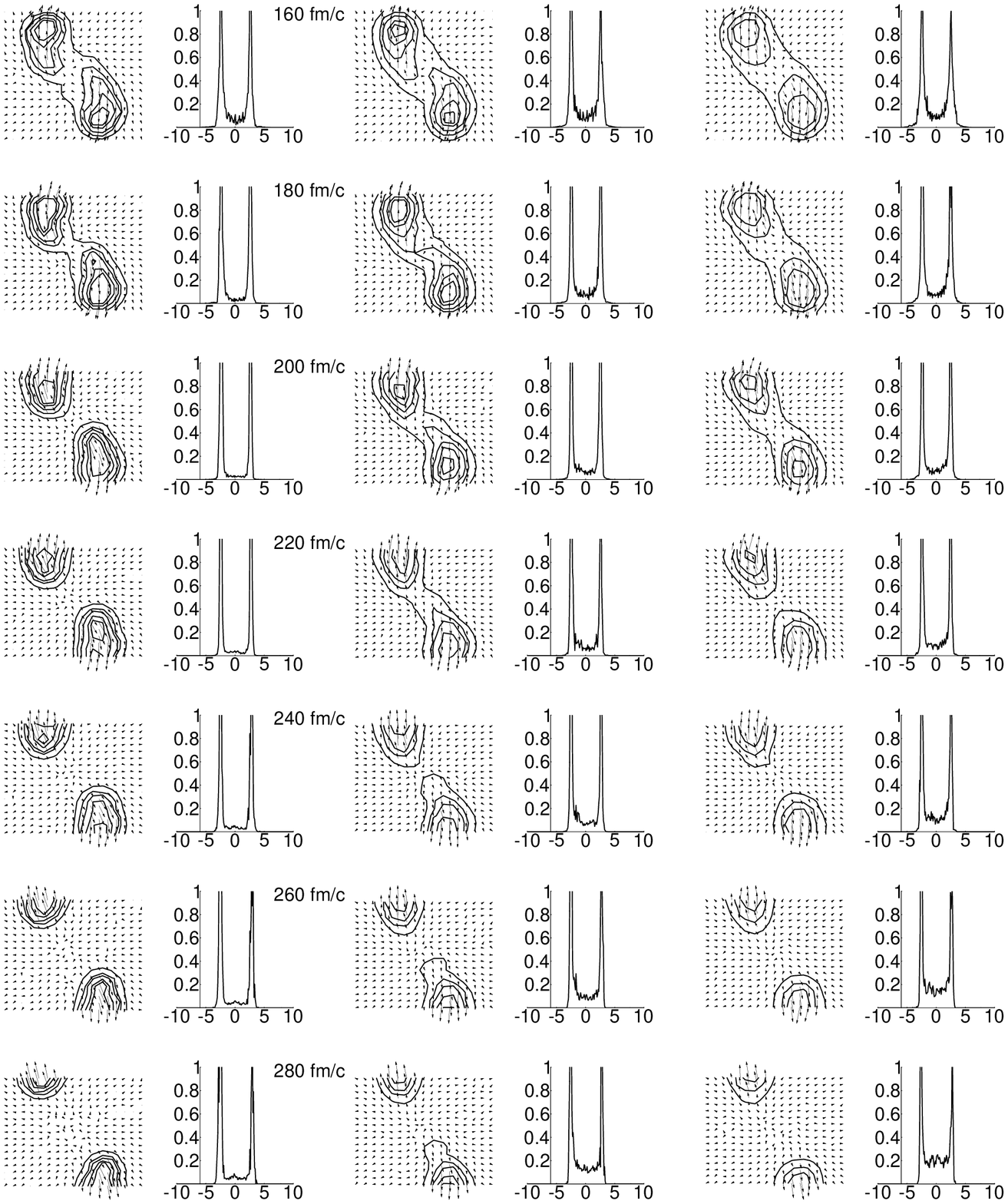,height=17cm,angle=0}
\caption{The time evolution of $^{181}_{73}Ta + ^{197}_{79}Au$ collisions at
$E_{lab}/A = 33$ MeV and $8$fm impact parameter
in  the BUU (left), nonlocal kinetic model (middle)
as well as the nonlocal model with quasiparticle
renormalisations (right). Plots
in the first column show the $(x-z)$ - density cut where $Ta$ as
projectile comes from below.
The mass momenta are shown by arrows. The corresponding second column
gives the charge density distribution versus relative velocity in
$cm/ns$ where the target like distribution of $Au$ is on the left and
the projectile like distributions of $Ta$ on the right.}
\label{midrap08}
\end{figure}

The evolution of the density can be seen in the corresponding left
pictures of figure~\ref{midrap08} for the
BUU (left panel) and nonlocal scenario (middle panel) as well as the additional quasiparticle
renormalisation (right panel). We see that the nonlocal scenario leads
to a longer and more pronounced neck formation between $200-240$fm/c
while the BUU breaks apart already at $200$fm/c.

The question arises whether this pronounced neck formation is simply
by more collisions and corresponding correlations. This would lead us
to the assumption that a simply increase in the cross section as
sometimes called in--medium effect would lead to the mid--rapidity
matter enhancement. This is however only the case for smaller impact
parameters \cite{G98}. To understand the
qualitative difference between the nonlocal scenario and an increase
of cross section by in--medium effects we perform a simulation where
in the local BUU scenario the cross section has been doubled. We see in
in the next figure \ref{n} the number of collisions per time for the
different scenarios.

\onecolumn
\twocolumn

\begin{figure}
\psfig{file=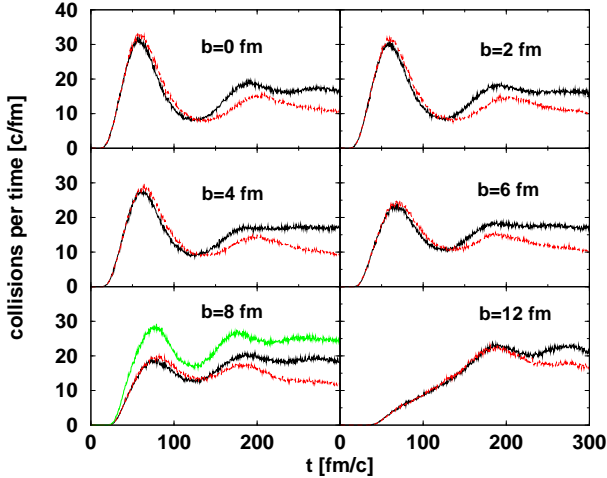,height=8cm,angle=-90}
\caption{The time evolution of the number of nucleon collisions for
  $Ta + Au$ at
$E_{lab}/A = 33$ MeV and different impact parameter in  the BUU (thick
black line), nonlocal kinetic model (broken line) and for the case of
$8$fm impact parameter the local BUU
with a cross section twice as large (thin dark line).}
\label{n}
\end{figure}

For $8$ fm impact parameter we compare the local
BUU (thick black line) with the nonlocal (broken line) and the local
BUU multiplying the cross section with two(thin line). We see that the
number of
collisions are visibly enhanced by doubling the cross section while
for the nonlocal scenario we get only a slight enhancement at the
beginning and later even lower values with respect to local BUU. The
latter fact comes from the earlier decomposition of matter in the
nonlocal scenario. Consequently from the number of collisions we would
conclude that the increase of cross section leads to more
correlations than the nonlocal scenario.

\begin{figure}
\psfig{file=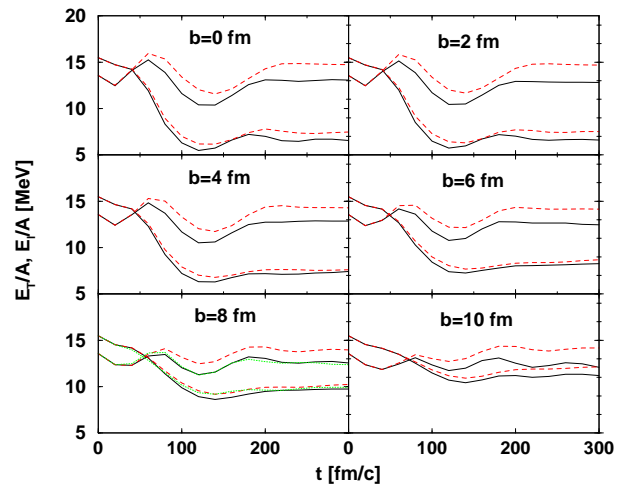,height=8cm,angle=-90}
\caption{The time evolution of the longitudinal (thin lines) and
  transverse energy (thick lines)
  including Fermi motion  of nucleon collisions for
  $Ta + Au$ at
$E_{lab}/A = 33$ MeV and different impact parameter in  the BUU (black
  line),
nonlocal kinetic equation (dashed line) and for the case of
$8$fm impact parameter the local BUU
but with twice the cross section (dotted line).}
\label{e}
\end{figure}

However when we look at the
corresponding transverse and kinetic energies in figure \ref{e}
($8$fm impact parameter) we
see that the transverse and longitudinal energy is almost not changed
compared with local BUU. Oppositely the nonlocal scenario leads to an
increase of transverse energy of about $2$MeV and about $1$MeV in
longitudinal energy. We conclude that the increase of cross section
leads to a higher number of collisions but not to more dissipated
energy while the nonlocal scenario does not change the number of
collisions much but the energy dissipated during the collisions. Roughly
speaking we can say that the quality of collision is changed.

Returning to the discussion of pronounced neck formation in figure
\ref{midrap08} above we see now that the quality
 rather than the quantity of collisions is what produces the neck. The simple
increase of the number of collision does not change much.

Now we can proceed and discuss the charge matter distribution with
respect to the velocity.
We evaluate the mean density and velocity, 
\be
n(r,t)&=&\int {d p\over (2\pi)^3} f(p,r,t)\nonumber\\
v(r,t)&=&\frac{1}{m \, n(r,t)} \int {d p\over (2\pi)^3} p f(p,r,t),
\ee
from which we define the distribution of hydrodynamical velocities
\be
F({\bar v},t)=\int dr\, n(r,t)\, \delta ({\bar v}-v_{\rm fiss}(r,t))
\ee
where $v_{\rm fiss}(r,t)$ is the projection of $v(r,t)$ onto the fission line.
This distribution we identify with the so called charge density distribution.

The definition of mean
mass (current) velocity does not include the Fermi energy which is
integrated out. In the case that we do have a different repartitioning
of Fermi energy during the collision than described in our kinetic
equation we will have here an ambiguity. Since the dynamical
cluster formation is not described in our approach we might have here
a smaller effect of Fermi energy on the mass velocity. This will lead us
indeed to the observation that BUU or nonlocal kinetic equations have
too much stopping compared to the experiment when more central
collisions are considered. For peripheral collisions we believe that
this kinetic description is sufficient which we will prove by proper
association of experimental events to the maximum in the  velocity
distribution.

We plot in figure \ref{midrap08}
also the normalised charge distribution versus velocity and see that
after $160$fm/c we have an appreciable higher mid--rapidity
distribution for the nonlocal scenario (mid panel) than the BUU
(left panel). Together with the observation that for nonlocal scenario
we have a pronounced neck formation we see indeed that the neck
formation is accompanied with high mid-velocity distribution of
matter.

\subsection{Quasiparticle renormalisation}
\begin{figure}
\psfig{file=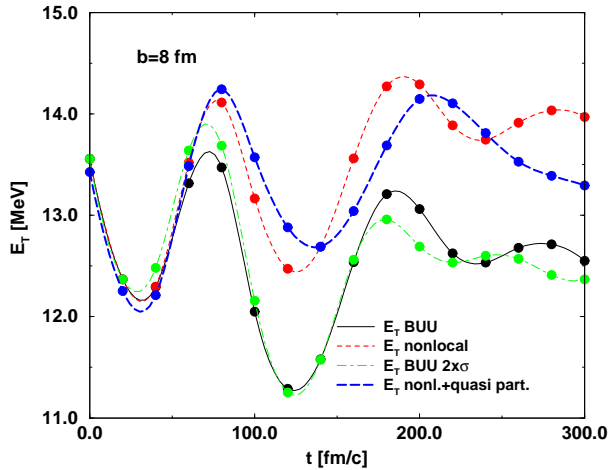,height=8cm,angle=-90}
\caption{The time evolution of the transverse energy
  including Fermi motion for
  $Ta + Au$ at
$E_{lab}/A = 33$ MeV and  $8$fm impact parameter in  the BUU (black
  line),
nonlocal kinetic equation (dashed line), the local BUU
but twice cross section (dashed dotted line) and the nonlocal scenario with
  quasiparticle renormalisation (long dashed line).}
\label{e08}
\end{figure}

Now we use the quasiparticle renormalisation schema which has been
outlined in chapter\ref{theoryren}. We see in figure \ref{midrap08}
(right panel) that the mid--rapidity distribution of matter is once more
enhanced in comparison to nonlocal scenario without quasiparticle
renormalisation. The seemingly shorter lifetime of the neck is
artificial due to
the chosen density contours which means we have lower densities and faster matter disintegration so that in fact
the neck is much more pronounced than in simple nonlocal scenario and
of course more pronounced than in BUU. The detailed comparison of the
time evolutions of the transverse energy for $8$fm impact parameter
can
be seen in figure \ref{e08}.

We recognise that the transverse energies including quasiparticle
renormalisation are
similar to the nonlocal scenario and higher than the BUU or BUU with
twice the cross section. However please remark that the
period of oscillation in the transverse energy which corresponds to a
giant resonance becomes larger for the case with quasiparticle
renormalisation. Since therefore the energy of this resonance
decreases we can conclude that the compressibility has been decreased
by the quasiparticle renormalisation. Sometimes this quasiparticle
renormalisation have been introduced by momentum dependent
mean-fields.
The effect is known to soften the
equation of state. We see here that we get a dynamical quasiparticle
renormalisation and a softening of equation of state.
This softening of equation of state is already slightly remarkable
when
the nonlocal
scenario is compared with BUU. With additional quasiparticle
renormalisation we see that this is much pronounced.

\subsection{Comparison with experiments}

\begin{figure}
\psfig{file=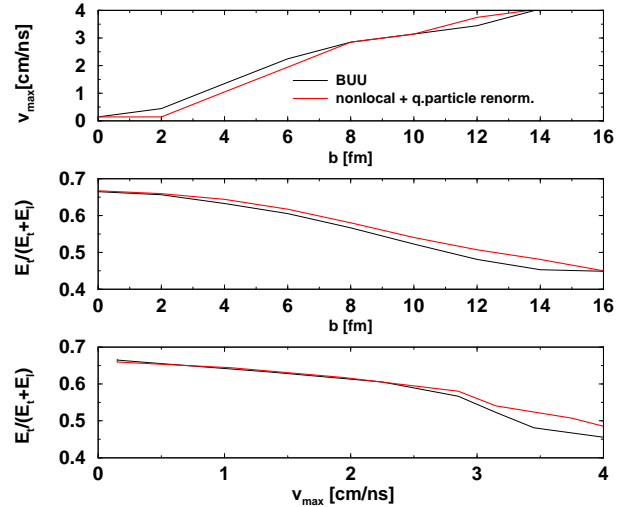,height=8cm,angle=-90}
\caption{The maximum velocity, impact parameter and ratio of
  longitudinal to total kinetic energy  of $Ta + Au$ collisions at
$E_{lab}/A = 33$ MeV
in  the BUU (solid line) and the nonlocal model with quasiparticle
renormalisation (dotted line).}
\label{betv}
\end{figure}

The BUU simulations will now be compared to one experiment performed
with INDRA at GANIL, the $Ta + Au$ collision at $E_{lab}/A = 33$ MeV.
The first question when comparing with experiments concerns the proper
selection of events such that one can compare with specific impact
parameter of the simulation. We choose here the point of view that the
maximum in the charge distribution with respect to velocity which is a
measure for stopping gives a good correlation with impact parameter.
Indeed if we compare the corresponding correlation between impact
parameter and this maximum velocity we obtain indeed an almost linear
correlation as in figure \ref{betv}.

\begin{figure}
\psfig{file=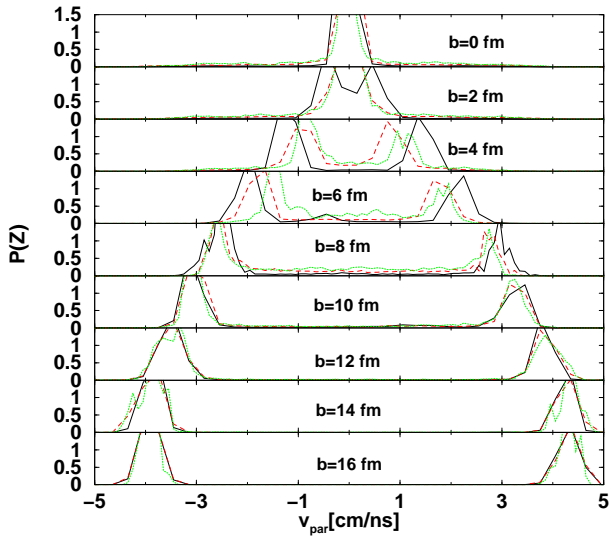,height=8cm,angle=-90}
\caption{The matter distribution versus velocity  of $Ta + Au$ collisions at
$E_{lab}/A = 33$ MeV and different impact parameter
in  the BUU (solid line), nonlocal kinetic equation (dashed line)
as well as the nonlocal model with quasiparticle
renormalisation (dotted line).}
\label{ch_n}
\end{figure}

The matter distribution is shown for different approximations in
figure \ref{ch_n}. One recognises clearly the successive enhancement of mid--rapidity matter around
$6-8$fm if one uses nonlocal kinetic theory and quasiparticle
renormalisation correspondingly.
It is interesting to remark that the dynamical quasiparticle
renormalisation which leads to a softening of the equation of state as
discussed in figure \ref{e08} enhances the mid rapidity distribution. In
contrast a mere soft static parametrisation of the mean-field does not
change the mid-rapidity emission appreciably \cite{G98}.

For the identification with experimental selection we use the
selection of events in the following way. First we select events which
show a clear one fragment structure. This correspond to events where
we have clear target and projectile like residues. Since the used
kinetic theory is not capable to describe dynamical fragment formation
we believe that these events are the one which are at least
describable within our frame. Next we use impact parameter cuts with
respect to the transverse energy since this shows in all simulation a
fairly good correlation to the impact parameter. In our numerical
results we see almost linear correlations between impact parameter,
maximal velocity and the convenient ratio between transverse and
total kinetic energy as seen in figure \ref{betv}.

\begin{figure}
\psfig{file=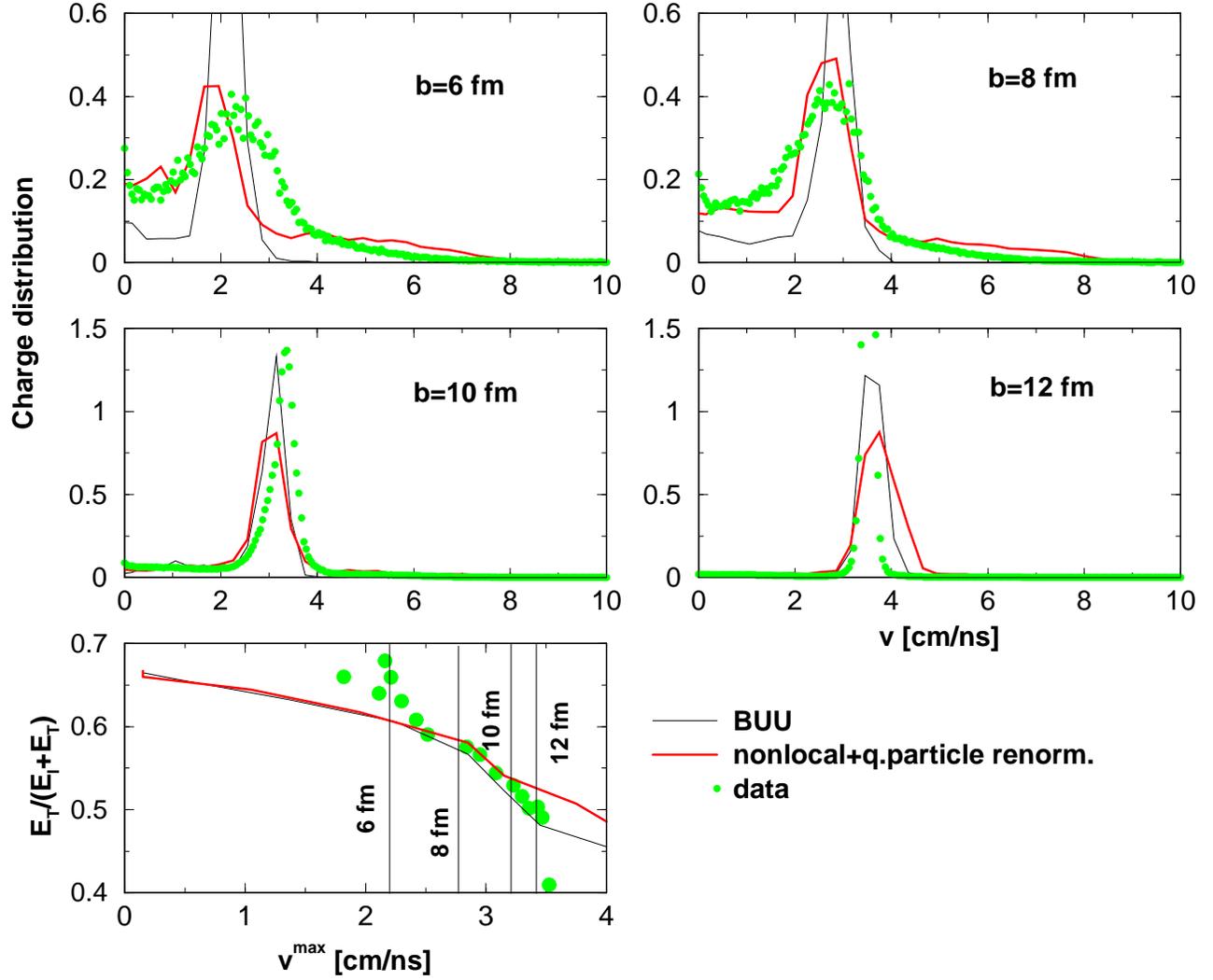,height=17cm,angle=-90}
\caption{The experimental charge distribution of matter (dotted line)
versus velocity in  comparison with in  the BUU (thin solid line) and
the nonlocal model with quasiparticle
renormalisation (thick line). The maximum velocity versus ratio of
  longitudinal to total kinetic energy  of $Ta + Au$ collisions at
$E_{lab}/A = 33$ MeV is given below. The selected experimental cuts are given by dots.}
\label{ch_p}
\end{figure}

For each selected experimental transverse energy bin we can plot now
the maximum velocity versus the ratio of the transverse to kinetic
energy. We see in figure \ref{ch_p} that the numerical velocity
damping agrees with the experimental selection
only for very peripheral collisions. For such events we plot in
the figure \ref{ch_p} the charge density distribution and compare the experiment with
the simulation. These charge density distributions have been obtained using the
procedure described in reference \cite{L00}. The Data are represented by light grey
points, the standard BUU calculation by the thin line and the non-local BUU
with quasiparticle renormalisation calculation by the thick line.  A
reasonable agreement is found for the nonlocal
scenario including quasiparticle renormalisation while simple BUU
fails to reproduce mid--rapidity matter.

\section{Summary and conclusion}

The extension of BUU simulations by nonlocal shifts and quasiparticle
renormalisation has been presented and compared to recent
experimental data on mid rapidity charge distributions. It is found
that both the nonlocal shifts as well as the quasiparticle renormalisation
must be included in order to get the observed mid--rapidity
matter enhancement.

The inclusion of quasiparticle renormalisation has been performed 
by using the normally  excluded events by Pauli
blocking. Since the quasiparticle renormalisation and corresponding
effective mass features can be considered as zero angle collisions
they can be realized by nonlocal shifts for the scattering events
which are normally rejected. This means that one has to perform the
advection step for the cases of Pauli blocked collisions without
colliding the particles. Besides giving a better description of experiments,
this has the effect of a dynamically softening of equation of state seen
in longer oscillations of giant compressional resonance.

In this way we present a combined picture including nonlocal off-sets
representing the nonlocal character of scattering, which leads to
virial correlations with the quasiparticle renormalisation, and as a
result to mean field fluctuations. We propose that no additional
stochasticity need to be assumed in order to get realistic
fluctuations.

\section{Acknowledgements}
The authors would like to thank the members of the INDRA collaboration for
providing the experimental data. N. Shannon is thanked for reading the
manuscript.

\onecolumn
\twocolumn


\end{document}